\documentclass[11pt,twoside]{article}  % Leave intact
\usepackage{adassconf}

\begin{document}   % Leave intact

%-----------------------------------------------------------------------
%			    Paper ID Code
%-----------------------------------------------------------------------
% Enter the proper paper identification code.  The ID code for your
% paper is the session number associated with your presentation as
% published in the official conference proceedings.  You can           
% find this number locating your abstract in the printed proceedings
% that you received at the meeting or on-line at the conference web
% site; the ID code is the letter/number sequence proceeding the title 
% of your presentation. 
%
% This will not appear in your paper; however, it allows different
% papers in the proceedings to cross-reference each other.  Note that
% you should only have one \paperID, and it should not include a
% trailing period.
%
% EXAMPLE: \paperID{O4-1}
% EXAMPLE: \paperID{P7-7}
%

\paperID{P3.22}

%-----------------------------------------------------------------------
%		            Paper Title 
%-----------------------------------------------------------------------
% Enter the title of the paper.
%
% EXAMPLE: \title{A Breakthrough in Astronomical Software Development}
% 
% If your title is so long as to fill the page header when you print it,
% then please supply a short form as a \titlemark.
%
% EXAMPLE: 
%  \title{Rapid Development for Distributed Computing, with Implications
%         for the Virtual Observatory}
%  \titlemark{Rapid Development for Distributed Computing}
%

\title{ASPID-SR: Prototype of a VO-compliant Science-Ready Data Archive}
\titlemark{ASPID-SR: a Prototype of VO-compliant Archive}

%-----------------------------------------------------------------------
%		          Authors of Paper
%-----------------------------------------------------------------------
% Enter the authors followed by their affiliations.  The \author and
% \affil commands may appear multiple times as necessary (see example
% below).  List each author by giving the first name or initials first
% followed by the last name.  Authors with the same affiliations
% should grouped together. 
%
% EXAMPLE: \author{Raymond Plante, Doug Roberts, 
%                  R.\ M.\ Crutcher\altaffilmark{1}}
%          \affil{National Center for Supercomputing Applications, 
%                 University of Illinois Urbana-Champaign, Urbana, IL
%                 61801}
%          \author{Tom Troland}
%          \affil{University of Kentucky}
%
%          \altaffiltext{1}{Astronomy Department, UIUC}
%
% In this example, the first three authors, "Plante", "Roberts", and
% "Crutcher" are affiliated with "NCSA".  "Crutcher" has an alternate 
% affiliation with the "Astronomy Department".  The fourth author,
% "Troland", is affiliated with "University of Kentucky"

\author{Igor Chilingarian \altaffilmark{1,2}}
\author{Victor Afanasiev \altaffilmark{3}}
\author{Francois Bonnarel \altaffilmark{4}}
\author{Serguei Dodonov \altaffilmark{3}}
\author{Mireille Louys \altaffilmark{4}}
\author{Ivan Zolotukhin \altaffilmark{1}}

\altaffiltext{1}{Sternberg Astronomical Institute, Moscow State University,
13 Universitetski prospect, Moscow, 119992, Russia}
\altaffiltext{2}{Centre de Recherche Astronomique de Lyon, Observatoire de Lyon; CNRS, UMR 5574; Universit\'e Claude Bernard Lyon-1;
Ecole Normale Sup\'erieure de Lyon, Lyon, France}
\altaffiltext{3}{Special Astrophysical Observatory, Russian Academy of Sciences}
\altaffiltext{4}{Centre de Donn\'ees Astronomiques de Starsbourg, Observatoire de Starsbourg; CNRS, UMR 7550; 
Universit\'e Louis Pasteur, Starsbourg, France}

%-----------------------------------------------------------------------
%			 Contact Information
%-----------------------------------------------------------------------
% This information will not appear in the paper but will be used by
% the editors in case you need to be contacted concerning your
% submission.  Enter your name as the contact along with your email
% address.
% 
% EXAMPLE:  \contact{Dennis Crabtree}
%           \email{crabtree@cfht.hawaii.edu}
%

\contact{Igor Chilingarian}
\email{chil@sai.msu.su}

%-----------------------------------------------------------------------
%		      Author Index Specification
%-----------------------------------------------------------------------
% Specify how each author name should appear in the author index.  The 
% \paindex{ } should be used to indicate the primary author, and the
% \aindex for all other co-authors.  You MUST use the following
% syntax: 
%
% SYNTAX:  \aindex{Lastname, F. M.}
% 
% where F is the first initial and M is the second initial (if
% used).  This guarantees that authors that appear in multiple papers
% will appear only once in the author index.  
%
% EXAMPLE: \paindex{Crabtree, D.}
%          \aindex{Manset, N.}        
%          \aindex{Veillet, C.}        
%
% NOTE: this information is also used to build the author list that
% appears in the table of contents.  Authors will be listed in the order
% of the \paindex and \aindex commmands.
%

\paindex{Chilingarian, I.}
\aindex{Bonnarel, F.}     % Remove this line if there is only one author
\aindex{Afanasiev, V.}
\aindex{Dodonov, S.}
\aindex{Louys, M.}
\aindex{Zolotukhin, I.}

%-----------------------------------------------------------------------
%		      Author list for page header	
%-----------------------------------------------------------------------
% Please supply a list of author last names for the page header. in
% one of these formats:
%
% EXAMPLES:
% \authormark{Lastname}
% \authormark{Lastname1 \& Lastname2}
% \authormark{Lastname1, Lastname2, ... \& LastnameN}
% \authormark{Lastname et al.}
%
% Use the "et al." form in the case of seven or more authors, or if
% the preferred form is too long to fit in the header.

\authormark{Chilingarian et al.}

%-----------------------------------------------------------------------
%			Subject Index keywords
%-----------------------------------------------------------------------
% Enter a comma separated list of up to 6 keywords describing your
% paper.  These will NOT be printed as part of your paper; however,
% they will be used to generate the subject index for the proceedings.
% There is no standard list; however, you can consult the indices
% for past proceedings (http://adass.org/adass/proceedings/).
%
% EXAMPLE:  \keywords{visualization, astronomy: radio, parallel
%                     computing, AIPS++, Galactic Center}
%
% In this example, the author noticed that "radio astronomy" appeared
% in the ADASS VII Index as "astronomy" being the major keyword and
% "radio" as the minor keyword.  The colon is used to introduce another
% level into the index.


%-----------------------------------------------------------------------
%			       Abstract
%-----------------------------------------------------------------------
% Type abstract in the space below.  Consult the User Guide and Latex
% Information file for a list of supported macros (e.g. for typesetting 
% special symbols). Do not leave a blank line between \begin{abstract} 
% and the start of your text.

\begin{abstract}          % Leave intact
ASPID stands for the "Archive of Spectral, Photometric, and Interferometric
Data".  The world largest collection of raw 3D spectroscopic observations of
galactic and extragalactic sources is provided. ASPID-SR is a prototype of
an archive of heterogeneous science ready data, fed by ASPID, where we try
to exploit all the power of the IVOA Characterisation Data Model.
Multi-level Characterisation metadata is provided for every dataset. The
archive provides powerful metadata query mechanism with access to every data
model element, vital for the efficient scientific usage of a complex
informational system. We provide a set of access interfaces: SIAP/SSAP,
HTTP-based characterisation metadata query, Web-service accepting ADQL/x. 
\end{abstract}

%-----------------------------------------------------------------------
%			      Main Body
%-----------------------------------------------------------------------
% Place the text for the main body of the paper here.  You should use
% the \section command to label the various sections; use of
% \subsection is optional.  Significant words in section titles should
% be capitalized.  Sections and subsections will be numbered
% automatically. 
%
% EXAMPLE:  \section{Introduction}
%           ...
%           \subsection{Our View of the World}
%           ...
%           \section{A New Approach}
%
% It is recommended that you look at the sample papers, sample1.tex
% and sample2.tex, for examples for formatting references, footnotes,
% figures, equations, html links, lists, and other special features.  

\section{Introduction}
Interoperability between various data archives and data discovery, retrieval
and analysis tools is a cornerstone for a success of the International
Virtual Observatory. Data model, providing full and self-sufficient
description of a dataset, is an essential for the interoperability.
Characterisation DM of IVOA (McDowell et al. 2006) is one of the most
general data models, positioning the dataset in the space of physical
parameters.

ASPID stands for the "Archive of Spectral, Photometric, and Interferometric
Data". It is operated by the Laboratory of Spectroscopy and Photometry of
Extragalactic Objects and contains over 72000 raw observational datasets
coming from selected instruments at the 6-m telescope of SAO RAS since 1992:
direct images; 1D, long-slit, multi-object, and IFU spectroscopy;
Fabry-Perot datacubes; spectropolarimetric observations. The world largest
collection of 3D spectroscopic observations of galactic and extragalactic
sources is provided.

\section{Implementation}
As a backend of ASPID-SR we use the open-source PostgreSQL database engine
with the initial XMLType support, implemented in a frame of the ''Google
Summer of Code-2006'' by Nikolai Samokhvalov (see poster by Zolotukhin et al.
for technical details of the database and query interface implementations).
Queries on spatial coordinates are implemented using pgSphere extension
(Chilingarian et al. 2004b), providing capabilities to deal with geometrical
objects on a sphere.

\section{Capabilities}
Multi-level Characterisation metadata is provided for every dataset. We have
shown applicability of Characterisation DM to description of 3D datasets
(Chilingarian et al. 2006). We followed the latest IVOA Data Models 
working group recommendations for mandatory Characterisation
metadata. For every dataset we provide first two graining levels of the
characterisation metadata (location or reference value and bounds) for
Coverage, Resolution, and Sampling Precision properties of Spatial,
Spectral, Time, and Observable (Flux) axes of the data model.

Queries will be allowed on all these metadata (including Characterisation
AxisFrame fields). This is a vital point for the efficient scientific usage
of such a complex informational system.

ASPID-SR archive is accessible via PHP-based HTTP-interface at \\
{\bf http://alcor.sao.ru/php/aspid-sr/}. Direct links to the datasets are
provided.

We will provide a set of VO-compliant access interfaces as well:
\begin{itemize}
\item SIAP/SSAP
\item HTTP-based characterisation metadata query
\item Web-service accepting ADQL/x
\end{itemize}

\section{Contents of ASPID-SR}
ASPID-SR provides reduced versions of raw datasets from the
ASPID archive, and source-result connection between two archives is
preserved. Only data older than the PI's proprietary period of two years
(according to SAO RAS Observational Data Archive Regulations), or datasets
provided by their PI are available in our science-ready data archive. Data
reduction of complex IFU spectra and IFP datasets was mostly done by members
of LSPEO: VA, SD, and Dr. Alexei Moiseev. Other data such as long-slit
spectra and direct images are normally reduced by PI's of the observing
programmes.

\section{Summary}
ASPID-SR is one of the first implementations of the Characterisation Data
Model, and we believe it is an important step toward development of VO-aware
complex astronomy informational systems.

\acknowledgments
We are very grateful to the support given by the organizing committee of
ADASS, essential for attending this exciting conference. Travel of IC is
supported via RFBR grant 06-02-27336

\end{document}